\documentclass[useAMS,usenatbib]{mn2e}
\usepackage{natbib}
\usepackage{graphicx}
\usepackage{amssymb}
\usepackage{caption}

\begin{document}

\newcommand{\cgs}{ ${\rm erg~cm}^{-2}~{\rm s}^{-1}$ }
\newcommand{\lum}{\rm erg~s$^{-1}$ }
\newcommand{\ha}{\rm H$\alpha$ }
\newcommand{\lo}{\rm $\lambda$ Ori }
\newcommand{\s}{\rm [SII]/\ha }
\newcommand{\n}{\rm [NII]/\ha }
\newcommand{\ox}{\rm [OIII]/\ha }
\newcommand{\he}{\rm [HeI]/\ha }
\newcommand{\apj}{\rm ApJ}
\newcommand{\mnras}{\rm MNRAS}
\newcommand{\pasa}{\rm PASA}
\newcommand{\araa}{\rm ARAA}
\newcommand{\apjs}{\rm ApJS}
\newcommand{\apjl}{\rm ApJL}
\newcommand{\aj}{\rm AJ}
\newcommand{\pasp}{\rm PASP}
\newcommand{\aap}{\rm AAP}

\title[Photoionisation and Heating of the ISM]{Photoionisation and Heating of a Supernova Driven, Turbulent Interstellar Medium}
\author[Barnes et al.]{J.E.~Barnes$^{1}$\thanks{email: jb652@st-andrews.ac.uk}, Kenneth Wood$^1$,  
Alex S. Hill$^2$, L.M.~Haffner$^{3,4}$
\newauthor 
\\
$^1$School of Physics \& Astronomy, University of St Andrews, North Haugh,
St Andrews, Fife, KY16 9SS, Scotland\\
$^2$ CSIRO Astronomy \& Space Science, Marsfield, NSW, Australia\\
$^3$Department of Astronomy, University of Wisconsin
Madison, 475 North Charter Street, Madison\\
$^4$  Space Science Institute, 4750 Walnut Street, Suite 205, Boulder, CO 80301}

\maketitle

\begin{abstract}

The Diffuse Ionised Gas (DIG) in galaxies traces photoionisation feedback from massive stars.  Through three dimensional photoionisation simulations, we study the propagation of ionising photons, photoionisation heating and the resulting distribution of ionised and neutral gas within snapshots of magnetohydrodynamic simulations of a supernova driven turbulent interstellar medium.  We also investigate the impact of non-photoionisation heating on observed optical emission line ratios.  
Inclusion of a heating term which scales less steeply with electron density than photoionisation is required to produce diagnostic emission line ratios similar to those observed with the Wisconsin \ha Mapper.  Once such heating terms have been included, we are also able to produce temperatures similar to those inferred from observations of the DIG, with temperatures increasing to above $15000$ K at heights $|z| \gtrsim 1$ kpc.  We find that ionising photons travel through low density regions close to the midplane of the simulations, while travelling through diffuse low density regions at large heights.  The majority of photons travel small distances ($\lesssim 100$pc); however some travel kiloparsecs and ionise the DIG. 

\end{abstract}

\section{Introduction}
The interstellar medium (ISM)  is a vital component of the cycle of star formation and the evolution of galaxies.   The composition and dynamics of the ISM determine the formation of new stars in the Galaxy, while stars provide feedback through ionisation, outflows and supernovae \citep{MacLow2004}. In this paper we study the formation of widespread ionised gas as observed primarily through \ha in the Milky Way and other galaxies. This gas (reviewed by \citealt{Haffner2009}), commonly referred to as the Diffuse Ionised Gas (DIG) or Warm Ionised Medium (WIM),  is low density ($\la 0.1$~cm$^{-3}$), warm ($\sim8000$~K), consists of regions of nearly fully ionised hydrogen (\citealt{Hausen2002}) and has a scale height of $1-1.5$ kpc near the sun (\citealt{Savage2009}, \citealt{Haffner1999}, \citealt{Gaensler2008}). 
  
The most likely sources of the ionisation of the DIG in the Galaxy are O stars \citep{Reynolds1990}.  Photoionisation simulations of a smooth ISM are able to reproduce some of the observed properties of the DIG (e.g., \citealt{Wood2004b}, \citealt{Miller1993}).  However, to allow ionising photons from midplane OB stars to propagate to large distances, these models require the vertical column density of hydrogen to be lower than that inferred from HI 21 cm observations of the Galaxy.  Photoionisation simulations of a clumpy or fractal density structure show that the introduction of lower density paths in a three dimensional (3D) ISM allow photons to travel to large heights above the midplane (see for example figure 16 of  \citealt{Haffner2009}).
The most likely source of such clumping is turbulence (e.g.,  \citealt{Armstrong1995}, \citealt{Hill2008}, \citealt{Chepurnov2010}, \citealt{Burkhart2012}) which could be driven by supernovae (e.g., \citealt{MacLow2004}, \citealt{Avillez2000}, \citealt{Armstrong1995}).  \cite{Wood2010} demonstrated that in a 3D supernova-driven, turbulent medium, ionising photons are indeed able to propagate to large distances and produce the DIG.  However, the width of the distribution of emission measure in these simulations is wider than observed in the Galaxy. The discrepancy appears to be due to too wide a variation of density with height, requiring a mechanism to smooth out the density variations in the dynamical simulations.  One possible smoothing mechanism is pressure from magnetic fields. In this paper we extend the work of \cite{Wood2010} to study photoionisation of a supernova-driven, turbulent magnetised ISM, using the 3D density structures from the MHD simulations of \citet{Hill2012}. 
 

Our simulations naturally produce a vertically extended ionised component and a compact neutral layer of gas, in qualitative agreement with observations.  However the \ha intensity from the ionised layer has a smaller scale height than observed in the Galaxy. To better reproduce \ha observations in the Galaxy we have created models of a 3D fractal ISM which provide estimates for the density structure and column densities for future MHD simulations.  We  also investigate the distance travelled by ionising photons in the ISM and find that the majority of photons travel only short distances and only a small fraction need to travel kiloparsec distances to ionise the DIG.  

The outline of the paper is as follows: a description of observations of the DIG is given in section \ref{WHAM}.  The MHD and Monte Carlo photoionisation codes are briefly described in section \ref{models}, the results of our simulations are presented in section \ref{results} and are compared with observations of the DIG in the Galaxy.  In  section \ref{analytic} we describe the results of photoionisation models of a 3D fractal ISM. In section \ref{results:paths} we investigate how far photons are able to travel through the ISM to create the DIG.  Finally our conclusions are presented in section \ref{conclusions}.  

\section{Wisconsin H$\alpha$ Mapper}
\label{WHAM}

Although the DIG was first detected at radio frequencies \citep{Hoyle1963}, much of our knowledge of its properties come from observations of optical emission lines (see review by \citealt{Haffner2009}).  In our present study we utilise data from the Wisconsin \ha Mapper (WHAM) survey which has mapped the entire northern sky in \ha \citep{Haffner2003} and large sections in other optical emission lines including H$\beta$, [N~II] $\lambda 6584${\AA} and [S~II] $\lambda 6716${\AA} (\citealt{Haffner1999}, \citealt{Madsen2005}, \citealt{Madsen2006}), and targeted observations of [O~I] $\lambda 6300${\AA}  (\citealt{Reynolds1998}, \citealt{Hausen2002}) and [O~III] $\lambda 5007${\AA}  \citep{Madsen2005}. These observations provide information on the distribution, kinematics, and physical conditions in the DIG (\citealt{Haffner1999}, \citealt{Haffner2003}).  Some of the main results from WHAM are that the temperature of the gas appears to increase with height from the midplane, inferred from the increase of [N~II]/H$\alpha$ and [S~II]/H$\alpha$ with height.  Similar trends in line ratio and temperature have also been observed in other galaxies (e.g., \citealt{Rand1998}, \citealt{Otte2002}).

We take advantage of the kinematic information provided by WHAM to compare our photoionisation simulations with observations of regions of the Galaxy with different star formation rates, namely gas associated with the Perseus Arm and an inter-arm region. The gas associated with the Perseus Arm is taken to be in the velocity range  $-75 $km s$^{-1}<V_{lsr}< -45$ km s$^{-1}$ and Galactic coordinate range $l=125^{\circ}$ to $156^{\circ}$ and $b=-6^{\circ}$ to $-35^{\circ}$.  There is some uncertainty as to the velocity of \ha  associated with the Perseus Arm with other authors taking the arm to be in the range $-50 $km s$^{-1}<V_{lsr}< -30$ km s$^{-1}$ \citep{Haffner1999}.  We use an inter-arm region in the Solar neighbourhood which lies in the velocity range $-15\,{\rm km\,s}^{-1} < V_{lsr}< 15\, {\rm km\, s}^{-1}$ and Galactic coordinate range $l=120^{\circ}$ to $160^{\circ}$ and $b=-30^{\circ}$ to $30^{\circ}$ (\citealt{Madsen2006}).  These regions were observed as part of the WHAM Northern Sky Survey and are seen clearly in \ha emission \citep{Madsen2006}.  In addition to H$\alpha$, maps of these regions in the [N II] $\lambda 6584${\AA} and [S II] $\lambda 6716${\AA} lines provide information on the temperature and ionisation state of the gas up to 2~kpc above the midplane of the Galaxy (\citealt{Haffner1999}, \citealt{Madsen2006}, \citealt{Madsen2005}).

\section{Models}
\label{models}

In this section we briefly describe the MHD and Monte Carlo photoionisation codes used in our theoretical study of diffuse ionised gas in the Galaxy.

\subsection{Magnetohydrodynamic Simulations}
\label{models:MHD}
We investigate the structure of DIG photoionised by OB stars using a 3D density structure from supernova driven MHD simulations of the ISM \citep{Hill2012b}.  These simulations  use FLASH v2.5 \citep{Fryxell2000} and are based on those of \cite{Joung2006} and \cite{Joung2009}.  A brief description of the simulations follows.

The density grid we use is from the magnetised ``bx50hr" simulation described by \cite{Hill2012} which employs an adaptive mesh grid with maximum resolution of 2~pc near the midplane and lower resolution at $|z|\gtrsim 50-300$pc.  The ``bx50hr'' model has an initial uniform horizontal magnetic field of 6.5$\mu$G in the midplane which approaches an rms value of $5-6 \mu$G (see review by \citet{Kulsrud2008}).  The full simulation grid is 1~kpc $\times$ 1~kpc $\times$ 40~kpc, with the midplane situated at the centre of the box. 

The MHD simulations use a modified version of the \cite{Kuijken1989} gravitational potential which includes a stellar disk, spherical dark matter halo, and a \cite{Navarro1996} profile above $|z|=8$ kpc. Supernova explosions drive turbulence  and structure the ISM, while heating and cooling establish a multiphase, vertically stratified ISM.  The supernovae are set off at approximately the Galactic supernova rate used by \cite{Joung2009}: Type Ia$=6.58~{\rm Myr}^{-1}{\rm kpc}^{-2}$, core collapse$=27.4~{\rm Myr}^{-1}{\rm kpc}^{-2}$.  Three fifths of the core collapse supernovae are clustered spatially and temporally to simulate super bubbles, though supernova positions and times are determined without knowledge of the gas distribution.  For our photoionisation simulation, we consider a snapshot at $t=316$~Myr. 

These simulations include a diffuse heating term representing photoelectric heating by dust grains \citep{Wolfire1995}.  They also contain heating terms from supernova explosions and stellar winds. Photoionisation is not considered and therefore heating by this mechanism is not included.  Radiative cooling is incorporated as appropriate for an optically thin, solar metallicity plasma in collisional ionisation equilibrium. 
  
  Since these simulations do not include photoionisation, they are unable to distinguish between the warm ionised and neutral medium.  The thermal pressure for a given density and temperature in the DIG is $\approx 2$ times that in the warm neutral medium due to the extra electron associated with each H atom.  However the impact of this on the dynamics should be small because in $10^{4}$ K gas the thermal pressure is significantly smaller than the turbulent pressure.  Additionally, simulations of high mass stars have shown that ionising radiation is able to produce molecular outflows.  However the properties of the outflows suggest that ionising radiation is not the main driver \citep{Peters2012}.

\begin{figure}
\begin{center}
\includegraphics[scale=0.5,trim=20mm 120mm 0mm 90 mm,clip]{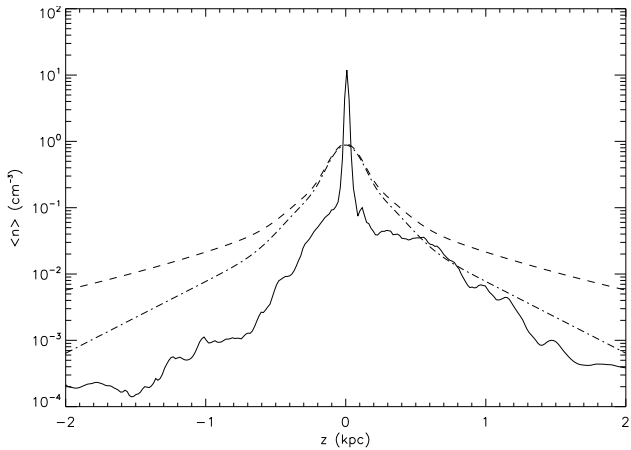}
\caption{Mean density (solid line) from MHD simulations, Dickey-Lockman density (dot-dash line) and Dickey-Lockman plus vertically extended gas component (dashed line).  The MHD simulation has densities $\approx$ 100 times lower at $|z|$ = 2kpc and is much more sharply peaked than the average density structure required to produce the observed scale height of H$\alpha$ in the Galaxy.}
\label{fig1}
\end{center}
\end{figure}

 Figure \ref{fig1} shows horizontally averaged density from the MHD simulation as a function of height ($z$).  We also show a Dickey-Lockman distribution for neutral hydrogen in the Galaxy \citep{DL1990} and a Dickey-Lockman plus vertically extended component, as required to match the observed H$\alpha$ in the Galaxy (\citealt{Reynolds1999}). Although the average vertical column density for the MHD models ($5.1 \times 10^{24}$cm$^{-2}$) is similar to that of the Dickey-Lockman distribution ($4.7\times10^{24}$cm$^{-2}$), the MHD density is more centrally peaked and is considerably lower at large $|z|$. As we will discuss in section \ref{results}, the MHD density structure affects the scale height of H$\alpha$ emission, temperature, and the line ratios \n and \s from the photoionisation models.  

\subsection{Photoionisation Models}
\label{models:photo}
In order to determine how ionising photons propagate and ionise the DIG we use the 3D Monte-Carlo photoionisation code of \cite{Wood2004}.  This model follows ionising photons through a 3D cartesian grid and computes the temperature and ionisation state of H, He, C, S, N, O and Ne in each cell.  Ionisation from both direct stellar and diffuse photons from recombinations of H$^+$ and He$^+$ are included.  Previous studies (e.g.,\citealt{Wood2010}) indicate that the impact of dust on photoionisation in the low density DIG is minimal and has therefore been neglected. We adopt the following abundances appropriate for the diffuse ISM: He/H$=0.1$, C/H$=1.4\times 10^{-4}$, 
N/H$=6.5\times 10^{-5}$, O/H$=4.3\times 10^{-4}$, Ne/H$=1.17\times 10^{-4}$, and S/H$=1.4\times 10^{-5}$.  The O/H and N/H abundances are averages from \cite{Jenkins2009} and \cite{Simpson2004}, S/H is taken from \cite{Daflon2009}, He/H and Ne/H are from \cite{Mathis2000}. 

 For the photoionisation simulations we use a section of the MHD grid that extends $\pm$2 kpc from the midplane. Memory restrictions of our photoionisation code on a desktop computer require us to rebin the density grid and reduce the resolution to 15.6 pc per grid cell. We have explored higher spatial resolution ionisation simulations for small sections of the grid as well as binning according to density-squared (because of the recombination rate dependence). The resulting local ionisation and temperature structures are unchanged at the 5\% level compared to the lower resolution runs.
 This resolution does not allow us to study the small scale ionisation of individual H{\sc ii} regions within the larger ISM grid.  Therefore the
 sources of ionising radiation in our simulation represent radiation escaping from H{\sc ii} regions into the large scale ISM and can be thought of as ``leaky H{\sc ii} regions" (e.g.,\citealt{Zurita2000}, \citealt{Zurita2002}).  The sources are randomly distributed in the $xy$ plane, and their $z$ location is randomly sampled from a gaussian distribution with a scale height of 63 pc, the observed scale height of O stars in the Galaxy \citep{Maiz2001}.  Heating is from photoionisation of the gas while cooling comes from recombination, free-free radiation and collisionally excited line radiation from C, N, O, Ne, and S. 
  We ignore the temperature structure from the MHD simulations, taking the density grid and calculating the temperature of the gas based only on heating by photoionisation plus an additional heating term (as required to explain emission line ratios in the gas).


We randomly place 24 sources in our simulations, following the study by \citet{Garmany1982} who estimated the surface density of O stars in the solar neighbourhood to be 24 stars kpc$^{-2}$.  The sources are distributed uniformly in x and y, with z chosen to reproduce the scale height of O stars in the Galaxy \citep{Maiz2001}.  
 The stars within 2.5~kpc of the Sun have a total estimated Lyman continuum luminosity $Q = 7 \times 10^{51}$ s$^{-1}$  \citep{Garmany1982}, while the total ionising luminosity of the stars in the Galaxy is estimated to be $2.6 \times 10^{53}$ s$^{-1}$ \citep{Williams97}.  Some ionising photons will produce the DIG, some will escape the Galaxy altogether, and the remainder will be trapped and produce local HII regions around each source.  As mentioned earlier, the small scale H{\sc ii} regions are not studied in this paper. We equally distribute the ionising luminosity among the 24 sources and investigate total luminosities escaping from HII regions $0.5 < Q_{49}<10$ where $Q_{49} \equiv Q/(10^{49} {\rm s}^{-1})$. 
 As expected, the ionising luminosity required to produce the DIG in our simulations is smaller than the total budget estimated by \cite{Vacca1996}  of $Q_{49} \approx 35$ for all stars in 1 kpc$^2$ of the disk. Like the dynamical simulations which produce the density grid, the radiation transfer simulation has repeating boundary conditions where photons that leave the simulation box from the $x$ or $y$ faces re-enter the box on the opposite side.  The ionising spectrum is taken from the library of radiation-driven wind atmosphere models for hot stars computed by \citet{Pauldrach2001} and \citet{Sternberg2003}.  
  The spectrum used for these simulations is from a model atmosphere with solar abundance, $\log {g} =3.4$ and $T=35000$~K.

For an input density structure from the MHD simulations, we calculate the ionisation and temperature structure arising from pure photoionisation. We do not include photoelectric heating or shock heating, the two heating effects included in the MHD simulations. Therefore the very high temperatures ($\sim 10^{6}$K) in some regions of the MHD simulations are not accounted for in the photoionisation models.
Such regions represent the hot ionised medium component of the ISM (\citealt{Mckee77}, \citealt{Cox2005}) and have a volume filling factor of around 60\% in the sub-grid from the MHD simulation (i.e., $|z| < 2$~kpc), in good agreement with other observational and theoretical studies of the ISM (e.g., \citealt{Mckee77}, \citealt{Harfst2006}). Since H$\alpha$ emissivity is proportional to $T^{-0.9}$ \citep{agn2} and the gas is at very low density, the very hot regions will produce very low H$\alpha$ intensity and can be ignored when computing the total intensity maps from the photoionisation models. We have produced H$\alpha$ maps with and without emission from the hot cells from the MHD simulation; they are almost identical for the reasons we have outlined, however the low \ha intensity and high temperature of these cells makes analysis of our results more difficult. 
Therefore, in the following analysis the intensity and line ratio maps from the photoionisation models do not include emission from the regions in the MHD simulations that have $T>30000$~K.

\section{Results}
\label{results}

\begin{figure*}
\begin{center}
\includegraphics{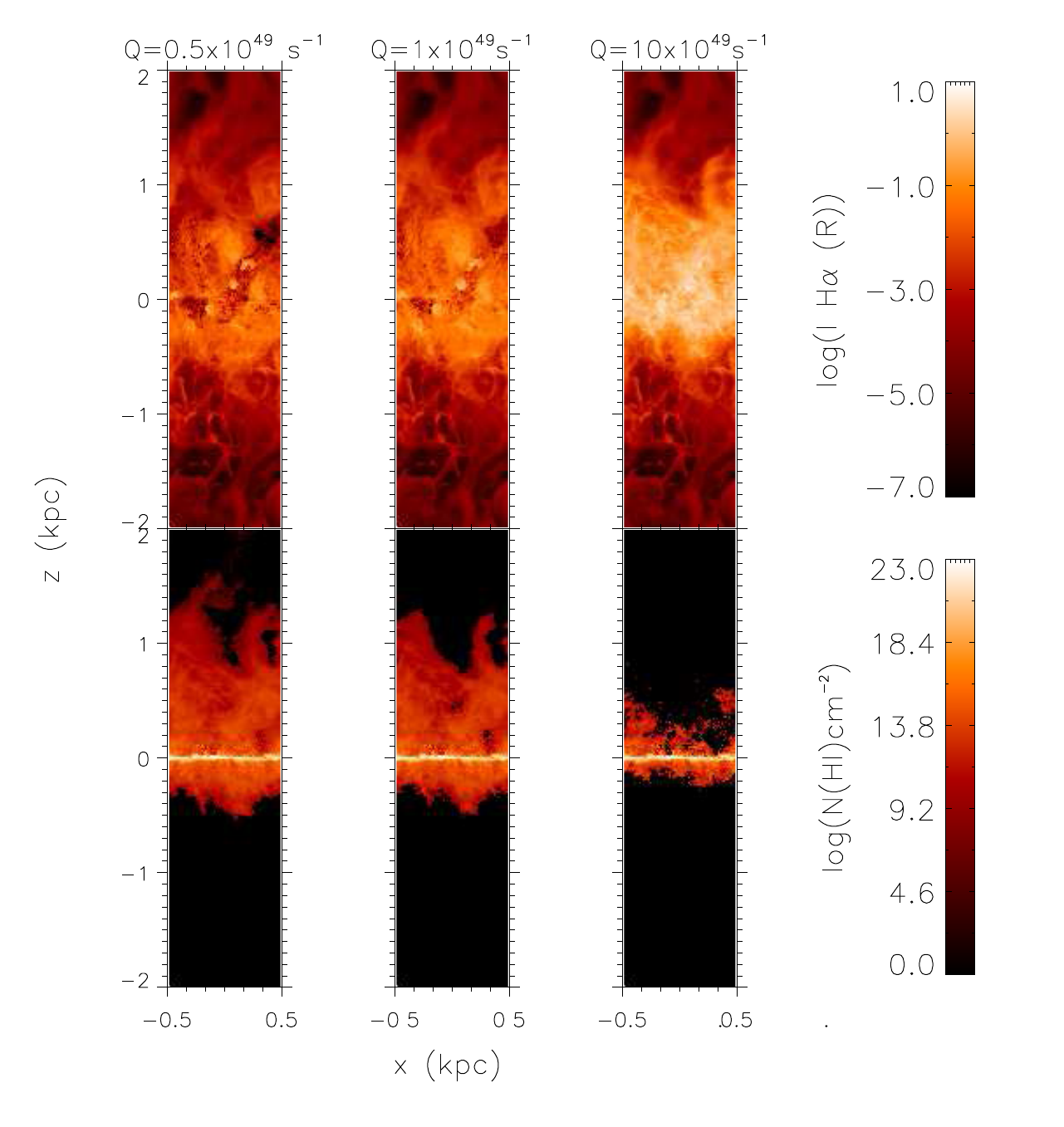}
\caption{Edge on view of H$\alpha$ intensity (top row) and column density of neutral hydrogen (bottom row) shown for increasing ionising luminosity, Q$_{49}$ = 0.5 (left), 1 (centre), 10 (right).}
\label{fig2}
\end{center}
\end{figure*}

 Figure \ref{fig2} shows maps of \ha intensity and column density of neutral hydrogen for different ionising luminosities in our simulations. As in \cite{Wood2010}, the photoionisation simulations naturally produce ionised gas at all heights and a less vertically extended distribution of neutral hydrogen. As the ionising luminosity is increased, the vertical extent of the neutral hydrogen decreases. These general properties are in qualitative agreement with observations in the Milky Way and many other galaxies. 

\citet{Reynolds1990} estimated that approximately 12\% of the ionising photons from OB stars are required to support the ionisation of the DIG. We find that a luminosity of $1\la Q\la 10 \times 10^{49}$s$^{-1}$ is able to ionise hydrogen to heights of $\pm$2~kpc and produce a compact distribution of neutral gas. This implies that for our simulations, around 3\% to 28\% of the estimated Lyman continuum budget of $Q_{49} \approx 35$ for 1kpc$^2$ \citep{Vacca1996} is required to escape from H{\sc ii} regions to ionise the DIG. 
 Similarly to
\cite{Wood2010}, we find that the most important parameter is the ionising luminosity and that the position of the sources has little effect on ionisation of the gas at large $|z|$.  A fraction of the photons in our simulations, typically less than a few percent, escape the simulation grid altogether (i.e. beyond $z=\pm 2$~kpc) and represent the escape fraction of ionising photons into the intergalactic medium. Recent galaxy-wide simulations estimate the escape fraction of ionising photons from dwarf galaxies into the IGM to be in the range 0.08\% and 5.9\% \citep{Kim2013}, while \cite{Barger2013} find that up to 4\% of ionising photons escape from the SMC while up to 5.5\% escape from the LMC, encompassing the values from our study.

\begin{figure}
\begin{center}
\includegraphics[scale=0.5,trim = 25mm 120mm 0mm 90 mm, clip]{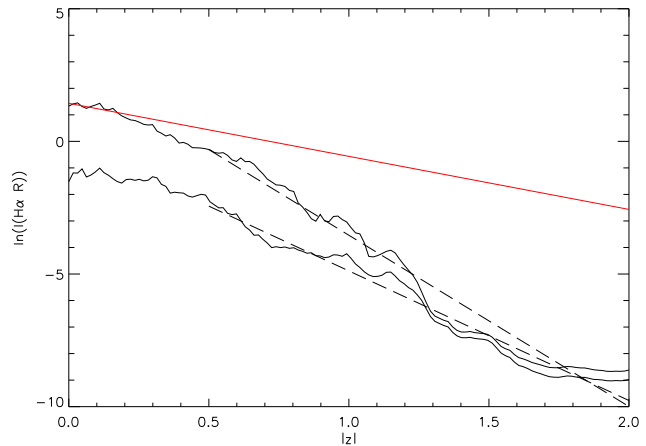}
\caption{Fit to the horizontally averaged \ha intensity vs height observed in the Perseus Arm (red), compared with photoionisation models from figure \ref{fig2}.  Top line: $Q_{49} = 10$ and a fit to the region of the model corresponding to observations of the Perseus Arm (dashed line); bottom line $Q_{49} = 1$  and fit (dashed line).  The scale height of \ha intensity for the Perseus arm is 500 pc. The simulations have scale heights of 150~pc ($Q_{49} = 10$) and 250~pc ($Q_{49}=1$). The H$\alpha$ intensity close to the midplane is small for $Q_{49}=1$ because this low ionising luminosity produces a smaller fraction of ionised gas close to the dense midplane regions compared to models with higher ionising luminosity.} 
\label{fig3}
\end{center}
\end{figure}

Figure \ref{fig3} shows that the scale height of \ha intensity from the MHD model is smaller than observed in the Milky Way.  For simulations with $Q_{49} = 10$, the \ha scale height is 150~pc while for $Q_{49} = 1$ it is 250~pc. These \ha scale heights are smaller than the typical 500~pc observed in the Milky Way because of the lower scale height of the input density grid. Compared to low-$Q$ models, in the high-$Q$ models there is more ionised gas and hence a larger \ha intensity at small $|z|$ and a smaller \ha scale height (see figures \ref{fig2} and \ref{fig4}). 

\subsection{Emission Line Ratios in the DIG}
 \label{results:lines} 
 
Observations of \s and \n in the DIG show these line ratios increase with decreasing H$\alpha$ intensity \citep{Haffner1999} while [SII]/[NII] remains almost constant.  It has been suggested that variations in temperature could explain the observed line ratios (e.g.,\citealt{BlandHawthorn1997}), with a higher temperature increasing \s and \n while keeping [SII]/[NII]  constant as result of the similar excitation potentials of S and N \citep{Reynolds1999}. The observed increase of \s and \n with altitude \citep{Haffner2009} suggests a temperature in the DIG that increases with $|z|$ and requires 
 additional physical processes that heat the gas in addition to photoionisation.  To provide additional heating in the low density gas at high $|z|$, these sources of heating should have a shallower dependence on density than the $n_e^2$ dependence of photoionisation heating.
 Possible mechanisms include heating from the dissipation of turbulence (proportional to $n_e$, \citealt{Minter1997}) and cosmic ray heating (proportional to $n_e^{-1/2}$, \citealt{Wiener2013}). 

Following the nomenclature in previous papers (\citealt{Reynolds1999}, \citealt{Wiener2013}, \citealt{Wood2004b}), the heating/cooling equation can be written,

\begin{equation}
G_0\, n_e^2 + G_1\, n_e + G_3\, n_e^{-\frac{1}{2}} = \Lambda\, n_e^2\; ,
\end{equation}
where the heating from photoionisation ($G_0$) and the cooling function ($\Lambda$) are computed explicitly in our code. For these simulations, we introduce one additional heating term $G_1\, n_e$ ergs cm$^{-3}$ s$^{-1}$, which will dominate over photoionisation heating at low values of $n_e$. In the analytic fractal density structures considered in section \ref{analytic} we also consider additional heating from cosmic rays with the term $G_3\, n_e^{-1/2}$ ergs cm$^{-3}$ s$^{-1}$.

\begin{figure}
\begin{center}
\includegraphics[scale=0.5,trim = 25mm 120mm 0mm 90 mm, clip]{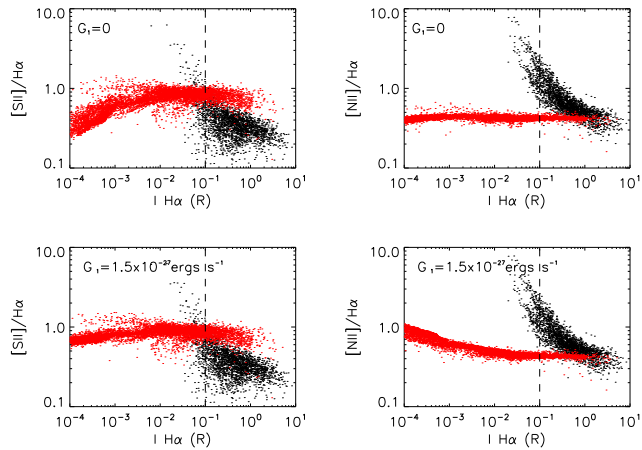}
\caption{Emission line ratios showing \n versus H$\alpha$ intensity for MHD simulations with $Q_{49} =1$ (red points) compared with WHAM data from the Perseus Arm (black points). The vertical line represents the WHAM sensitivity limit.  Upper and lower panels show simulations without and with additional heating respectively.} 
\label{fig4}
\end{center}
\end{figure}

\begin{figure}
\begin{center}
\includegraphics[scale=0.5,trim = 25mm 120mm 0mm 90 mm, clip]{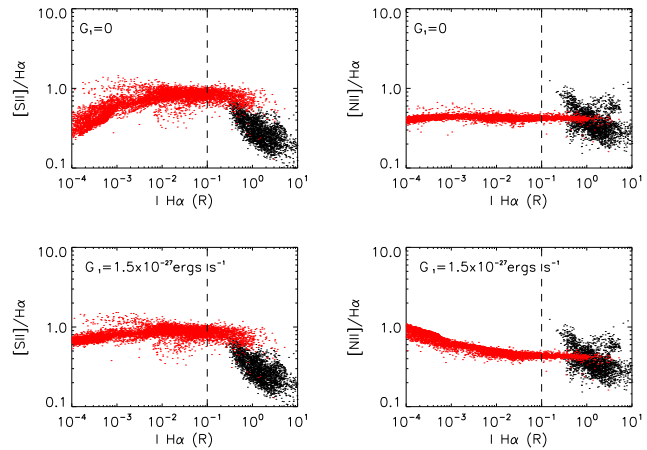}
\caption{Emission line ratios showing \n and \s versus H$\alpha$ intensity for MHD simulations with $Q_{49} =1$(red points) compared with WHAM data from the inter arm region (black points).  The vertical line represents the WHAM sensitivity limit.  Upper and lower panels show simulations without and with additional heating respectively.} \label{fig5}
\end{center}
\end{figure}

Figures \ref{fig4} and \ref{fig5} show diagnostic line ratio plots of \s and \n versus \ha from our simulations without (top panels) and with (bottom panels) additional heating $G_1\,n_e$. In Figure \ref{fig4} we compare the models to observations of the Perseus Arm, restricting the models to heights $|z| < 1.8$~kpc.  Figure \ref{fig5} compares our models to observations of the lower star formation rate inter-arm region. Without additional heating, the \s and \n line ratios show little variation with H$\alpha$ intensity and certainly no indication of the observed increase of the line ratios towards low H$\alpha$ intensities. Including an additional heating term with $G_{1}=1.5\times10^{-27}$ ergs~s$^{-1}$ does raise the [N~II]/H$\alpha$, but only at extremely small H$\alpha$ intensities.

 It is important to note that the MHD simulations were performed for the average galactic supernova rate from \cite{Joung2009} with type Ia and core collapse rates of $6.58$Myr$^{-1}$ kpc$^{-2}$ and $27.4$Myr$^{-1}$ kpc$^{-2}$.  The supernova rate is likely to be higher in the Perseus arm where stars are forming.  It is therefore appropriate to compare the simulated line ratios with an inter-arm region where the star formation and supernova rate will be closer to the average rates for the Galaxy.   Observations of \n and \s line ratios in the inter-arm region have a smaller increase with decreasing \ha intensity. However, due to the low \ha intensity at large $|z|$ in our simulations where the density is very small, we are still unable to match these ratios.
 There is a significant difference between the line ratios from regions below and above the midplane in the simulations and is attributed to the asymmetric density structure in the MHD simulation (see figure  \ref{fig1}).
 It is notable that we have assumed a path length through our simulations of $1$kpc, however this is an arbitrary choice.  If we adopt a path length of $2$kpc then the H$\alpha$ intensity would be twice as large, and therefore closer to that observed in the Perseus Arm, while leaving the line ratios unchanged.

 We have been unable to produce any increase in \s towards low H$\alpha$ intensities even with an additional heating term.
The difficulty in reproducing observations of S lines is not unexpected and is most likely because the dielectronic recombination rates for S are unknown. Most photoionisation codes either ignore dielectronic recombination for S or, as employed here, use averages of the rates for C, N, and O neglecting any temperature dependence (\citealt{Ali1991}).

 \subsection{Temperature}
\label{results:temps}
\begin{figure}
\begin{center}
\includegraphics[scale=0.5,trim = 25mm 120mm 20mm 90 mm, clip]{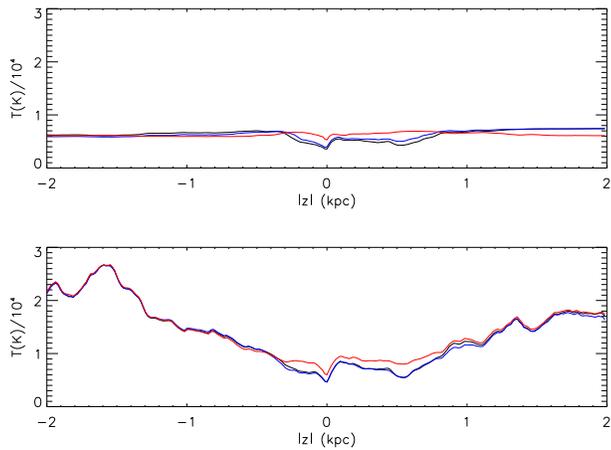}
\caption{Average vertical temperature structure of  photoionised gas in simulations with (top panel) no additional heating and (bottom panel) additional heating $G_{1}n_{e} = 1.5\times10^{-27}$ ergs~s$^{-1}n_{e}$ for ionising luminosities. Q$_{49}$ = 10 (red), 1(blue), 0.5 (black).}
\label{fig6}
\end{center}
\end{figure}

 With the assumption that N is primarily in the singly-ionised state throughout the DIG,
 observations of the line ratios in the Perseus Arm have been analysed to determine the temperature structure in the gas indicating that it rises from around 7000~K at $|z|=0.75$ kpc to over 10000 K above  $|z|=1.75$kpc (\citealt{Haffner1999}; \citealt{Madsen2006}). In figure \ref{fig6} we present the average temperature of ionised material in our simulations with and without the additional heating term, $G_1$.  To produce figure \ref{fig6} we have applied a cut where cells with temperature over 30000 K in the MHD simulations (typically corresponding to density below $\approx 5\times 10^{-4}$ cm$^{-3}$) have been neglected.  These cells represent parts of ``bubbles''  in the MHD simulations which in reality will be shock heated and collisionally ionised, processes that we do not consider in our current pure photoionisation models. 

Figure \ref{fig6} shows that without the addition of a non-photoionisation heating term, we are unable to reproduce the inferred increase in temperature with height.  However, once additional heating ($G_{1}n_{e}=1.5 \times10^{-27}\, n_e$ergs cm$^{-3}$s$^{-1}$) has been added, we do reproduce this increase.  We find that this increase is larger than inferred by \citet{Haffner1999}, with the gas temperature increasing to above 17000K at 1.75 kpc because of the very small densities for $|z|\ga 300$~pc in the MHD simulation. 
 A smaller value for $G_1$ gives lower temperatures but does not give a noticeable increase of \n at low $n_{e}$.
There is a significant difference between the line ratios from regions below and above the midplane in the simulations and is attributed to the asymmetric density structure in the MHD simulation (see figure  \ref{fig1}).

\section{Fractal Model of the ISM}
\label{analytic}

In the previous section we have shown that photoionisation models of a turbulent ISM can reproduce general trends in \ha emission observed in the Galactic DIG, 
 and that to increase emission line ratios an additional heating component is required.
 However, due to the small scale height in the MHD simulations, the scale height of \ha is smaller than observed in the Perseus Arm of the Galaxy.  
As a guide for future MHD simulations of the Galactic ISM, in this section we explore the photoionisation of analytic density structures that can better reproduce the observed average vertical distribution of H$\alpha$ intensity and emission line ratios observed by WHAM. 

 We adopt a smooth, four-component ISM density structure comprising a Dickey-Lockman distribution plus a more vertically extended component to produce the H$\alpha$ emission from the DIG. To allow ionising photons from midplane OB stars to propagate to large heights, we turn the smooth density into a 3D fractal structure using the fractal algorithm of \cite{Elmegreen1997} as described by \cite{Wood2005}. We adopt the same five-level hierarchical clumping algorithm, casting 16 seeds at the first level and 32 at each subsequent level. The density structure is arranged so that 33\% of the mass is smoothly distributed with the remaining fraction in the hierarchical clumps. For further details of the algorithm see figure~4 of \cite{Wood2005} and the accompanying description.

A density structure that produces the observed scale height of H$\alpha$ with an input ionising luminosity of $Q_{49}=16$ is

\begin{eqnarray}
\nonumber n(z) = 0.4{\rm e}^{-\left(|z|/h_{1}\right)^{2}/2}+0.11{\rm e}^{-\left(|z|/h_{2}\right)^{2}/2} \\
+0.06{\rm e}^{-|z|/h_{3}} + 0.04 {\rm e}^{-|z|/h_4}
\end{eqnarray}
where the height $z$ is measured in pc and the number densities are per cm$^3$.  The first three components represent the Dickey-Lockman distribution shown in figure \ref{fig1} with scale heights $h_1 =90$~pc, $h_2 = 225$~pc, $h_3 = 400$~pc. The fourth term represents the low density extended diffuse ionised gas and we take the scale height to be $h_4 = 1000$~pc, as typically inferred from H$\alpha$ observations in the Galaxy. 

 As before, we place 24 ionising sources in the simulation and assign each source a $T=35000$~K model atmosphere spectrum corresponding to an O5II star  \citep{Underhill1979}.  

\begin{figure}
\begin{center}
\includegraphics[scale=0.55,trim = 22mm 120mm 20mm 80 mm, clip]{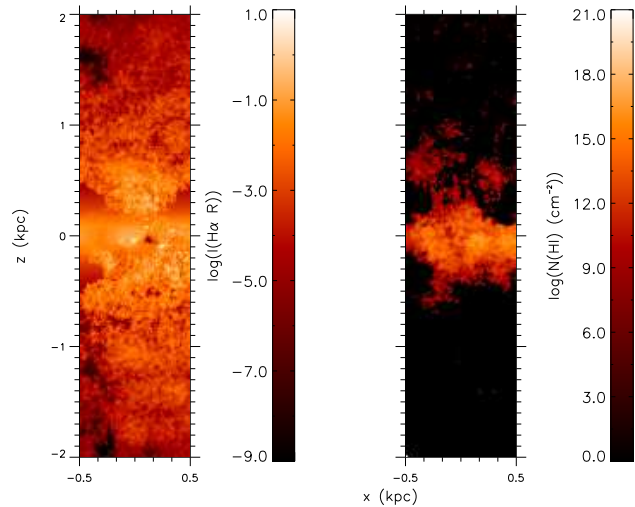}
\caption{  Edge on view of \ha intensity (left) and column density of neutral hydrogen (right) for an analytic fractal model of the DIG. }
\label{fig7}
\end{center}
\end{figure}

\begin{figure}
\begin{center}
\includegraphics[scale=0.5,trim = 22mm 120mm 20mm 80 mm, clip]{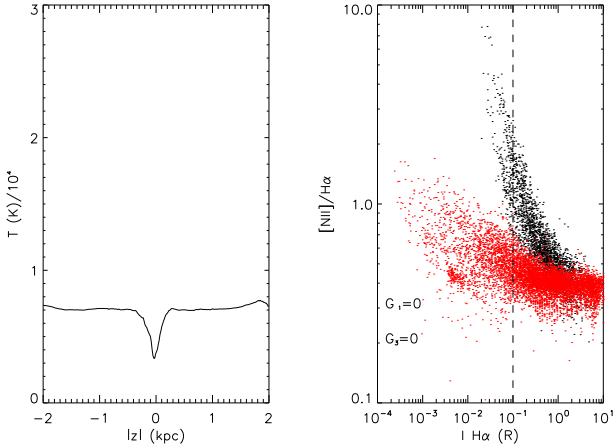}
\caption{Analytic fractal model of the DIG.  Left: average vertical temperature structure of ionised gas.   Right: \n vs \ha for the Perseus arm (black) and this model (red).  The vertical line represents the WHAM sensitivity limit. Here we show \n line ratios only in the region that corresponds to the Perseus Arm ($|z| < 1.8$ kpc).  Although this model seems able to reproduce the observed \ha and neutral H, \n and temperature are different to those expected, suggesting the need for an additional heating term to increase temperature and \n at large $|z|$.}
\label{fig8}
\end{center}
\end{figure}

\begin{figure}
\begin{center}
\includegraphics[scale=0.5,trim = 22mm 120mm 20mm 80 mm, clip]{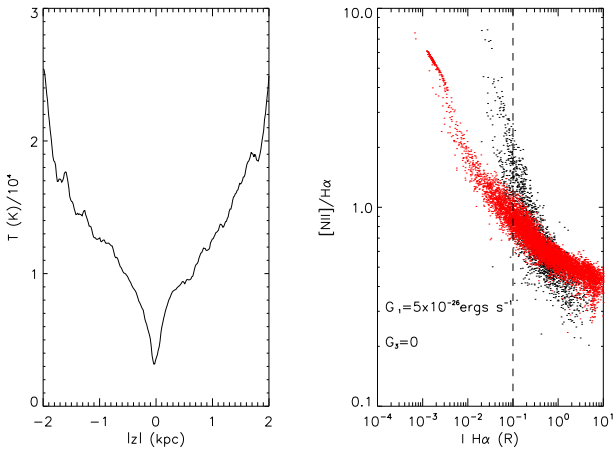}
\caption{Same as figure \ref{fig8} including additional heating $G_{1} = 4 \times 10^{26}$.  This model is now able to produce \n and temperatures similar to those inferred.  Here we show \n line ratios only in the region that corresponds to the Perseus Arm ($|z| < 1.8$ kpc).  }
\label{fig9}
\end{center}
\end{figure}

\begin{figure}
\begin{center}
\includegraphics[scale=0.5,trim = 22mm 120mm 20mm 80 mm, clip]{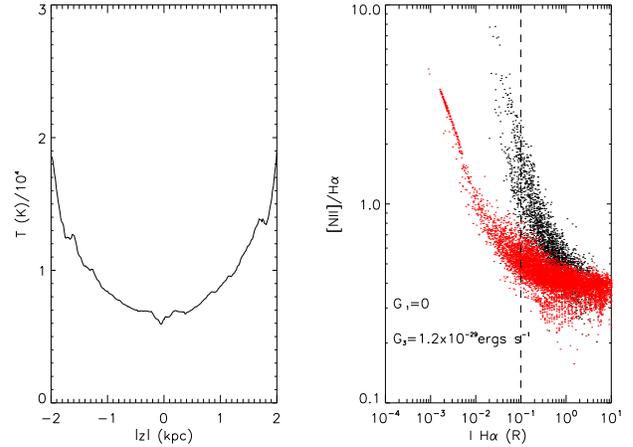}
\caption{Same as figure \ref{fig8} including additional heating from cosmic rays $G_{3}$ from equation \ref{cr_heat}. Here we show \n line ratios only in the region that corresponds to the Perseus Arm ($|z| < 1.8$ kpc.) }
\label{fig10}
\end{center}
\end{figure}

Figure \ref{fig7} shows the edge-on view of the H$\alpha$ intensity and the neutral hydrogen column density for the fractal density structure described above. The low density pathways of the interclump medium allow ionising photons to reach and ionise gas many kpc from the midplane and the H$\alpha$ intensity and \n line ratios shown in figures \ref{fig8}, \ref{fig9}, and \ref{fig10} are similar to observed in the Perseus Arm. This is not surprising as we have adopted a density structure that will produce the intensity distribution of H$\alpha$ in line with the analytic results presented by \cite{Haffner1999} and \cite{Reynolds1999}. While the fractal models, with additional heating terms, reproduce many of the features in diagnostic line ratio plots, the H$\alpha$ intensity map does not exhibit the filamentary structures, loops, and bubbles present in the MHD simulations. Therefore the fractal models should serve as a guide for the average density structure and additional heating required for future MHD simulations of the ISM and DIG.

The resulting average temperature structure and \n versus \ha are shown for simulations with no additional heating (figure \ref{fig8}), additional heating proportional to $n_e$ (figure \ref{fig9}) with $G_1 = 4\times10^{-26}$, and an additional heating term to simulate cosmic ray heating (figure \ref{fig10}) parameterised by
  \begin{equation}
  G_{3}=1.2\times10^{-29}\,e^{-3|z|/4000pc}\; {\rm erg}\,{\rm cm}^{-3}\,{\rm s}^{-1} , 
  \label{cr_heat}
  \end{equation} 
 as in equation 1.  This equation for cosmic ray heating was determined by \cite{Wiener2013} via analytic modelling of the [N II] and H$\alpha$ emission and assuming that N$^+$/H$^+$ is constant in the DIG and assuming a 1D density structure.  $G_{3}$ is one hundred times smaller than the value found by \cite{Wiener2013}, because our simulations determine the actual ionisation fractions, and include the spatial variation of ionisation, heating and cooling due to the 3D density, radiation field, and temperature structure.
Modelling data using our 3D photoionisation code therefore provides a better estimate of the additional heating terms required to reproduce the observed trends in the diagnostic diagrams.

\section{How Far Can Photons Travel to Ionise the DIG?}
\label{results:paths}

\begin{figure}
\begin{center}
                \includegraphics[scale=0.55,trim = 30mm 130mm 0mm 90 mm, clip]{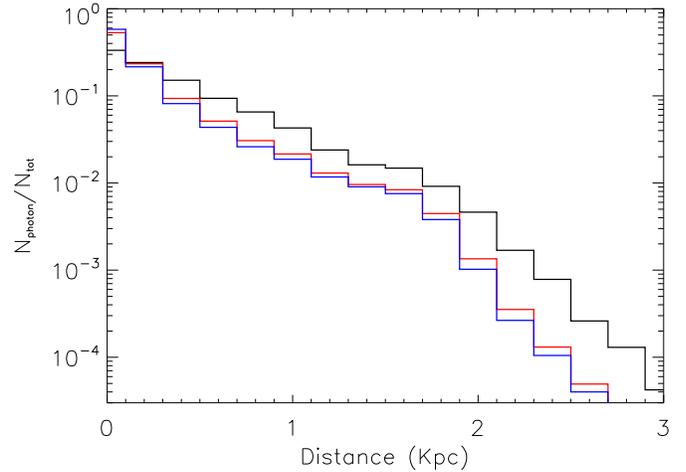}
                \caption{Histogram of distances traveled by ionising photons through the MHD simulation grid for ionising luminosities Q$_{49}$= 10 (black), 1 (red), 0.5 (blue). }
                \label{fig11}
\end{center}
\end{figure}

 How far photons are able to travel through the ISM and ionise the DIG is a question that is important to studies not only of DIG in galaxies, but also for the escape of ionising photons into the intergalactic medium. The 3D structures in the MHD and fractal density grids provide low density pathways allowing ionising photons to reach far above the midplane.
  Figure \ref{fig11} shows a histogram of the distances travelled by ionising photons in the density structure provided by the MHD simulations. Some photons do indeed travel very large distances from their sources and are responsible for ionising the gas at large $|z|$. The majority of the ionising photons ionise denser gas towards the midplane and only a small fraction is required to ionise the low density high altitude gas. Increasing the ionising luminosity allows photons to travel larger distances because more of the gas at low altitudes is ionised and hence presents a very small opacity.

\begin{figure}
\begin{center}
                \includegraphics[scale=0.55,trim = 30mm 130mm 0mm 90 mm, clip]{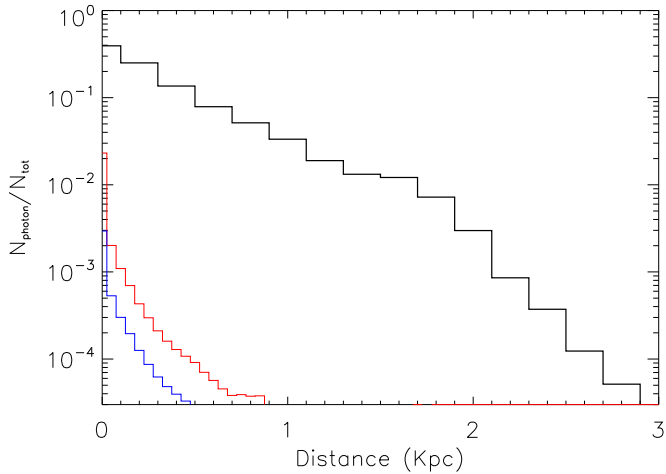}
\caption{Histogram showing distances traveled by all ionising photons (black), diffuse photons from hydrogen (red) and helium (blue) for Q$_{49}=1$.}
                \label{fig12}
\end{center}
\end{figure}
 
Our photoionisation code tracks direct stellar photons and also diffuse ionising radiation from H and He recombination. The diffuse ionising radiation consists of Lyman continuum from recombinations to the ground state of H and He, the He two-photon continuum, and the 19.8~eV line from He \citep{Wood2004}. Figure \ref{fig12} shows a histogram of the distances travelled by direct and diffuse ionising photons and shows that the majority of the diffuse photons are absorbed close to their location of emission (i.e., the ``on the spot" approximation in many photoionisation codes), but some do travel many hundreds of parsecs through the MHD density grid.

\begin{figure}
\begin{center}
\includegraphics [scale=0.55,trim = 15mm 120mm 0mm 80 mm, clip]{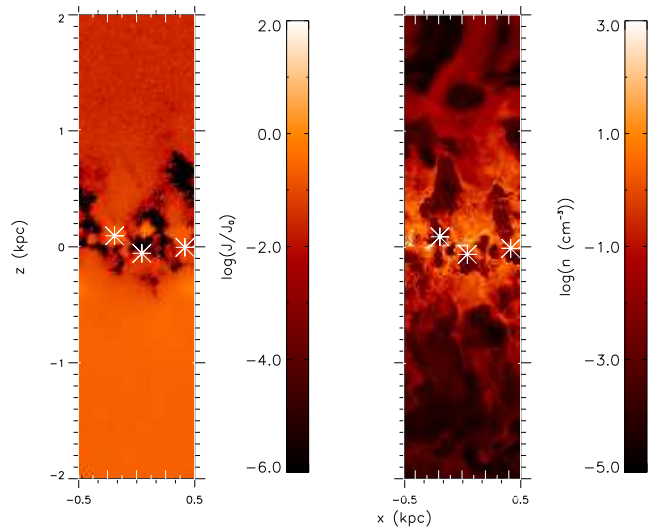}
\caption{Maps showing mean intensity of ionising radiation (left) and density (right) in one slice of the simulation box.  White stars indicate the position of ``bubbles'' in the MHD simulations and are characterised by low density regions in the density structure (right).  Bright regions of intensity show areas in which many photons are travelling while dark regions have few photons entering them.  There is a strong relation between the density of the region and the number of photons travelling in that gas with low density regions having many photons in them and high density regions appearing dark in mean intensity.  }
\label{fig13}
\end{center}
\end{figure}

Figure \ref{fig13} shows the mean intensity of the ionising photons in a one pixel wide slice of the simulation box compared to the density in that slice with positions of ``bubbles'' marked with a star.  Photons travel primarily in the low density regions of the grid at high altitude, but close to the midplane they travel primarily in bubbles or structures attached to these bubbles. The brightest regions in the mean intensity map correspond to source locations and in the case where a source has been randomly placed in a high density region close to the midplane, its ionising photons are trapped close to the source.  At higher altitudes the mean intensity becomes more uniform but higher density regions are still visible with fewer photons penetrating into and through these cells.  All of the ``bubbles'' in the simulations are located close to the midplane of the box and so at high altitudes photons appear to be travelling through a low density medium. This provides further evidence that compared to a smooth density structure a 3D ISM naturally allows for ionising photons to penetrate to larger distances and produce widespread diffuse ionised gas.

\section{Conclusions}
\label{conclusions}
 We have produced photoionisation simulations of the DIG in an environment similar to that in the outer disc of a spiral galaxy and compared these to observations of the Perseus Arm and an inter-arm region in the solar neighbourhood.  
We summarise our main conclusions here: 
\begin{enumerate}

\item   The photoionisation of 3D density structures from MHD simulations naturally produces widespread diffuse ionised gas with a density scale height larger than that of neutral hydrogen.  However the density grids from the MHD simulations we have used here have a low scale height  resulting in smaller \ha  intensity scale height than observed in the Perseus Arm.  

\item  We find that with the addition of a heating term, such as heating by cosmic rays or dissipation of turbulence, our simulations are able to reproduce general trends in \n  optical emission line ratios in the DIG. 

\item  A fractal density structure for the ISM with higher density at large $|z|$ than in the MHD simulations better reproduces WHAM observations of the Perseus arm.   This will provide a guide for the required density structure for future MHD simulations of the DIG. 

\item   Finally our simulations demonstrate that ionising photons are able to travel many kiloparsecs  to ionise the DIG at large heights above the midplane, with photons travelling through low density ``bubbles'' close to the midplane and through low density diffuse gas at large heights.

An important next step in modelling the ionisation of the DIG will be to include photoionisation as a dynamical process in MHD simulations of the gas (e.g., \citealt{deavillez2012}).  This may increase the density of gas at large heights since photoionisation will increase the temperature of the gas and allow the gas to expand to a larger height, thereby sustaining higher densities above the midplane of the Galaxy that are demanded by the WHAM observations.

\end{enumerate}

\section*{acknowledgments}
The Wisconsin H-Alpha Mapper is funded by the US National Science Foundation.
JB acknowledges the support of an STFC studentship.  The authors would like to thank Mordecai-Mark Mac Low for his helpful comments on this manuscript.

\bibliographystyle{mn2e}
\bibliography{mnraspaper}

\begin{thebibliography}{61}
\expandafter\ifx\csname natexlab\endcsname\relax\def\natexlab#1{#1}\fi

\bibitem[{{Ali} {et~al}\mbox{.}(1991){Ali}, {Blum}, {Bumgardner}, {Cranmer},
  {Ferland}, {Haefner}, \& {Tiede}}]{Ali1991}
{Ali} B., {Blum} R.~D., {Bumgardner} T.~E., {Cranmer} S.~R., {Ferland} G.~J.,
  {Haefner} R.~I., {Tiede} G.~P., 1991, \pasp, 103, 1182

\bibitem[{{Armstrong}, {Rickett} \& {Spangler}(1995){Armstrong}, {Rickett}, \&
  {Spangler}}]{Armstrong1995}
{Armstrong} J.~W., {Rickett} B.~J., {Spangler} S.~R., 1995, \apj, 443, 209

\bibitem[{{Barger}, {Haffner} \& {Bland-Hawthorn}(2013){Barger}, {Haffner}, \&
  {Bland-Hawthorn}}]{Barger2013}
{Barger} K.~A., {Haffner} L.~M., {Bland-Hawthorn} J., 2013, ArXiv e-prints

\bibitem[{{Bland-Hawthorn}, {Freeman} \& {Quinn}(1997){Bland-Hawthorn},
  {Freeman}, \& {Quinn}}]{BlandHawthorn1997}
{Bland-Hawthorn} J., {Freeman} K.~C., {Quinn} P.~J., 1997, \apj, 490, 143

\bibitem[{{Burkhart}, {Lazarian} \& {Gaensler}(2012){Burkhart}, {Lazarian}, \&
  {Gaensler}}]{Burkhart2012}
{Burkhart} B., {Lazarian} A., {Gaensler} B.~M., 2012, \apj, 749, 145

\bibitem[{{Chepurnov} \& {Lazarian}(2010)}]{Chepurnov2010}
{Chepurnov} A., {Lazarian} A., 2010, \apj, 710, 853

\bibitem[{{Cox}(2005)}]{Cox2005}
{Cox} D.~P., 2005, \araa, 43, 337

\bibitem[{{Daflon} {et~al}\mbox{.}(2009){Daflon}, {Cunha}, {de la Reza},
  {Holtzman}, \& {Chiappini}}]{Daflon2009}
{Daflon} S., {Cunha} K., {de la Reza} R., {Holtzman} J., {Chiappini} C., 2009,
  \aj, 138, 1577

\bibitem[{{de Avillez}(2000)}]{Avillez2000}
{de Avillez} M.~A., 2000, \mnras, 315, 479

\bibitem[{{de Avillez} {et~al}\mbox{.}(2012){de Avillez}, {Asgekar},
  {Breitschwerdt}, \& {Spitoni}}]{deavillez2012}
{de Avillez} M.~A., {Asgekar} A., {Breitschwerdt} D., {Spitoni} E., 2012,
  \mnras, 423, L107

\bibitem[{{Dickey} \& {Lockman}(1990)}]{DL1990}
{Dickey} J.~M., {Lockman} F.~J., 1990, \araa, 28, 215

\bibitem[{{Elmegreen}(1997)}]{Elmegreen1997}
{Elmegreen} B.~G., 1997, \apj, 477, 196

\bibitem[{{Fryxell} {et~al}\mbox{.}(2000){Fryxell}, {Olson}, {Ricker},
  {Timmes}, {Zingale}, {Lamb}, {MacNeice}, {Rosner}, {Truran}, \&
  {Tufo}}]{Fryxell2000}
{Fryxell} B. {et~al.}, 2000, \apjs, 131, 273

\bibitem[{{Gaensler} {et~al}\mbox{.}(2008){Gaensler}, {Madsen}, {Chatterjee},
  \& {Mao}}]{Gaensler2008}
{Gaensler} B.~M., {Madsen} G.~J., {Chatterjee} S., {Mao} S.~A., 2008, \pasa,
  25, 184

\bibitem[{{Garmany}, {Conti} \& {Chiosi}(1982){Garmany}, {Conti}, \&
  {Chiosi}}]{Garmany1982}
{Garmany} C.~D., {Conti} P.~S., {Chiosi} C., 1982, \apj, 263, 777

\bibitem[{{Haffner} {et~al}\mbox{.}(2009){Haffner}, {Dettmar}, {Beckman},
  {Wood}, {Slavin}, {Giammanco}, {Madsen}, {Zurita}, \&
  {Reynolds}}]{Haffner2009}
{Haffner} L.~M. {et~al.}, 2009, Reviews of Modern Physics, 81, 969

\bibitem[{{Haffner}, {Reynolds} \& {Tufte}(1999){Haffner}, {Reynolds}, \&
  {Tufte}}]{Haffner1999}
{Haffner} L.~M., {Reynolds} R.~J., {Tufte} S.~L., 1999, \apj, 523, 223

\bibitem[{{Haffner} {et~al}\mbox{.}(2003){Haffner}, {Reynolds}, {Tufte},
  {Madsen}, {Jaehnig}, \& {Percival}}]{Haffner2003}
{Haffner} L.~M., {Reynolds} R.~J., {Tufte} S.~L., {Madsen} G.~J., {Jaehnig}
  K.~P., {Percival} J.~W., 2003, \apjs, 149, 405

\bibitem[{{Harfst}, {Theis} \& {Hensler}(2006){Harfst}, {Theis}, \&
  {Hensler}}]{Harfst2006}
{Harfst} S., {Theis} C., {Hensler} G., 2006, \aap, 449, 509

\bibitem[{{Hausen} {et~al}\mbox{.}(2002){Hausen}, {Reynolds}, {Haffner}, \&
  {Tufte}}]{Hausen2002}
{Hausen} N.~R., {Reynolds} R.~J., {Haffner} L.~M., {Tufte} S.~L., 2002, \apj,
  565, 1060

\bibitem[{{Hill} {et~al}\mbox{.}(2008){Hill}, {Benjamin}, {Kowal}, {Reynolds},
  {Haffner}, \& {Lazarian}}]{Hill2008}
{Hill} A.~S., {Benjamin} R.~A., {Kowal} G., {Reynolds} R.~J., {Haffner} L.~M.,
  {Lazarian} A., 2008, \apj, 686, 363

\bibitem[{{Hill} {et~al}\mbox{.}(2012{\natexlab{a}}){Hill}, {Joung}, {Mac Low},
  {Benjamin}, {Haffner}, {Klingenberg}, \& {Waagan}}]{Hill2012b}
{Hill} A.~S., {Joung} M.~R., {Mac Low} M.-M., {Benjamin} R.~A., {Haffner}
  L.~M., {Klingenberg} C., {Waagan} K., 2012{\natexlab{a}}, \apj, 761, 189

\bibitem[{{Hill} {et~al}\mbox{.}(2012{\natexlab{b}}){Hill}, {Joung}, {Mac Low},
  {Benjamin}, {Haffner}, {Klingenberg}, \& {Waagan}}]{Hill2012}
{Hill} A.~S., {Joung} M.~R., {Mac Low} M.-M., {Benjamin} R.~A., {Haffner}
  L.~M., {Klingenberg} C., {Waagan} K., 2012{\natexlab{b}}, \apj, 750, 104

\bibitem[{{Hoyle} \& {Ellis}(1963)}]{Hoyle1963}
{Hoyle} F., {Ellis} G.~R.~A., 1963, Australian Journal of Physics, 16, 1

\bibitem[{{Jenkins}(2009)}]{Jenkins2009}
{Jenkins} E.~B., 2009, \apj, 700, 1299

\bibitem[{{Joung} \& {Mac Low}(2006)}]{Joung2006}
{Joung} M.~K.~R., {Mac Low} M.-M., 2006, \apj, 653, 1266

\bibitem[{{Joung}, {Mac Low} \& {Bryan}(2009){Joung}, {Mac Low}, \&
  {Bryan}}]{Joung2009}
{Joung} M.~R., {Mac Low} M.-M., {Bryan} G.~L., 2009, \apj, 704, 137

\bibitem[{{Kim} {et~al}\mbox{.}(2013){Kim}, {Krumholz}, {Wise}, {Turk},
  {Goldbaum}, \& {Abel}}]{Kim2013}
{Kim} J.-h., {Krumholz} M.~R., {Wise} J.~H., {Turk} M.~J., {Goldbaum} N.~J.,
  {Abel} T., 2013, \apj, 775, 109

\bibitem[{{Kuijken} \& {Gilmore}(1989)}]{Kuijken1989}
{Kuijken} K., {Gilmore} G., 1989, \mnras, 239, 605

\bibitem[{{Kulsrud} \& {Zweibel}(2008)}]{Kulsrud2008}
{Kulsrud} R.~M., {Zweibel} E.~G., 2008, Reports on Progress in Physics, 71,
  046901

\bibitem[{Mac~Low \& Klessen(2004)}]{MacLow2004}
Mac~Low M.-M., Klessen R.~S., 2004, Rev. Mod. Phys., 76, 125

\bibitem[{{Madsen} \& {Reynolds}(2005)}]{Madsen2005}
{Madsen} G.~J., {Reynolds} R.~J., 2005, \apj, 630, 925

\bibitem[{{Madsen}, {Reynolds} \& {Haffner}(2006){Madsen}, {Reynolds}, \&
  {Haffner}}]{Madsen2006}
{Madsen} G.~J., {Reynolds} R.~J., {Haffner} L.~M., 2006, \apj, 652, 401

\bibitem[{{Ma{\'{\i}}z-Apell{\'a}niz}(2001)}]{Maiz2001}
{Ma{\'{\i}}z-Apell{\'a}niz} J., 2001, \aj, 121, 2737

\bibitem[{{Mathis}(2000)}]{Mathis2000}
{Mathis} J.~S., 2000, \apj, 544, 347

\bibitem[{{McKee} \& {Ostriker}(1977)}]{Mckee77}
{McKee} C.~F., {Ostriker} J.~P., 1977, \apj, 218, 148

\bibitem[{{Miller} \& {Cox}(1993)}]{Miller1993}
{Miller}, III W.~W., {Cox} D.~P., 1993, \apj, 417, 579

\bibitem[{{Minter} \& {Spangler}(1997)}]{Minter1997}
{Minter} A.~H., {Spangler} S.~R., 1997, \apj, 485, 182

\bibitem[{{Navarro}, {Frenk} \& {White}(1996){Navarro}, {Frenk}, \&
  {White}}]{Navarro1996}
{Navarro} J.~F., {Frenk} C.~S., {White} S.~D.~M., 1996, \apj, 462, 563

\bibitem[{{Osterbrock} \& {Ferland}(2006)}]{agn2}
{Osterbrock} D.~E., {Ferland} G.~J., 2006, {Astrophysics of gaseous nebulae and
  active galactic nuclei}

\bibitem[{{Otte}, {Gallagher} \& {Reynolds}(2002){Otte}, {Gallagher}, \&
  {Reynolds}}]{Otte2002}
{Otte} B., {Gallagher}, III J.~S., {Reynolds} R.~J., 2002, \apj, 572, 823

\bibitem[{{Pauldrach}, {Hoffmann} \& {Lennon}(2001){Pauldrach}, {Hoffmann}, \&
  {Lennon}}]{Pauldrach2001}
{Pauldrach} A.~W.~A., {Hoffmann} T.~L., {Lennon} M., 2001, \aap, 375, 161

\bibitem[{{Peters} {et~al}\mbox{.}(2012){Peters}, {Klaassen}, {Mac Low},
  {Klessen}, \& {Banerjee}}]{Peters2012}
{Peters} T., {Klaassen} P.~D., {Mac Low} M.-M., {Klessen} R.~S., {Banerjee} R.,
  2012, \apj, 760, 91

\bibitem[{{Rand}(1998)}]{Rand1998}
{Rand} R.~J., 1998, \apj, 501, 137

\bibitem[{{Reynolds}(1990)}]{Reynolds1990}
{Reynolds} R.~J., 1990, \apjl, 349, L17

\bibitem[{{Reynolds}, {Haffner} \& {Tufte}(1999){Reynolds}, {Haffner}, \&
  {Tufte}}]{Reynolds1999}
{Reynolds} R.~J., {Haffner} L.~M., {Tufte} S.~L., 1999, \apjl, 525, L21

\bibitem[{{Reynolds} {et~al}\mbox{.}(1998){Reynolds}, {Hausen}, {Tufte}, \&
  {Haffner}}]{Reynolds1998}
{Reynolds} R.~J., {Hausen} N.~R., {Tufte} S.~L., {Haffner} L.~M., 1998, \apjl,
  494, L99

\bibitem[{{Savage} \& {Wakker}(2009)}]{Savage2009}
{Savage} B.~D., {Wakker} B.~P., 2009, \apj, 702, 1472

\bibitem[{{Simpson} {et~al}\mbox{.}(2004){Simpson}, {Rubin}, {Colgan},
  {Erickson}, \& {Haas}}]{Simpson2004}
{Simpson} J.~P., {Rubin} R.~H., {Colgan} S.~W.~J., {Erickson} E.~F., {Haas}
  M.~R., 2004, \apj, 611, 338

\bibitem[{{Sternberg}, {Hoffmann} \& {Pauldrach}(2003){Sternberg}, {Hoffmann},
  \& {Pauldrach}}]{Sternberg2003}
{Sternberg} A., {Hoffmann} T.~L., {Pauldrach} A.~W.~A., 2003, \apj, 599, 1333

\bibitem[{{Underhill} {et~al}\mbox{.}(1979){Underhill}, {Divan},
  {Prevot-Burnichon}, \& {Doazan}}]{Underhill1979}
{Underhill} A.~B., {Divan} L., {Prevot-Burnichon} M.~L., {Doazan} V., 1979, in
  IAU Symposium, Vol.~83, Mass Loss and Evolution of O-Type Stars, {Conti}
  P.~S., {De Loore} C.~W.~H., eds., pp. 103--107

\bibitem[{{Vacca}, {Garmany} \& {Shull}(1996){Vacca}, {Garmany}, \&
  {Shull}}]{Vacca1996}
{Vacca} W.~D., {Garmany} C.~D., {Shull} J.~M., 1996, \apj, 460, 914

\bibitem[{{Wiener}, {Zweibel} \& {Oh}(2013){Wiener}, {Zweibel}, \&
  {Oh}}]{Wiener2013}
{Wiener} J., {Zweibel} E.~G., {Oh} S.~P., 2013, ArXiv e-prints

\bibitem[{{Williams} \& {McKee}(1997)}]{Williams97}
{Williams} J.~P., {McKee} C.~F., 1997, \apj, 476, 166

\bibitem[{{Wolfire} {et~al}\mbox{.}(1995){Wolfire}, {Hollenbach}, {McKee},
  {Tielens}, \& {Bakes}}]{Wolfire1995}
{Wolfire} M.~G., {Hollenbach} D., {McKee} C.~F., {Tielens} A.~G.~G.~M., {Bakes}
  E.~L.~O., 1995, \apj, 443, 152

\bibitem[{{Wood} {et~al}\mbox{.}(2005){Wood}, {Haffner}, {Reynolds}, {Mathis},
  \& {Madsen}}]{Wood2005}
{Wood} K., {Haffner} L.~M., {Reynolds} R.~J., {Mathis} J.~S., {Madsen} G.,
  2005, \apj, 633, 295

\bibitem[{{Wood} {et~al}\mbox{.}(2010){Wood}, {Hill}, {Joung}, {Mac Low},
  {Benjamin}, {Haffner}, {Reynolds}, \& {Madsen}}]{Wood2010}
{Wood} K., {Hill} A.~S., {Joung} M.~R., {Mac Low} M.-M., {Benjamin} R.~A.,
  {Haffner} L.~M., {Reynolds} R.~J., {Madsen} G.~J., 2010, \apj, 721, 1397

\bibitem[{{Wood} \& {Mathis}(2004)}]{Wood2004b}
{Wood} K., {Mathis} J.~S., 2004, \mnras, 353, 1126

\bibitem[{{Wood}, {Mathis} \& {Ercolano}(2004){Wood}, {Mathis}, \&
  {Ercolano}}]{Wood2004}
{Wood} K., {Mathis} J.~S., {Ercolano} B., 2004, \mnras, 348, 1337

\bibitem[{{Zurita} {et~al}\mbox{.}(2002){Zurita}, {Beckman}, {Rozas}, \&
  {Ryder}}]{Zurita2002}
{Zurita} A., {Beckman} J.~E., {Rozas} M., {Ryder} S., 2002, \aap, 386, 801

\bibitem[{{Zurita}, {Rozas} \& {Beckman}(2000){Zurita}, {Rozas}, \&
  {Beckman}}]{Zurita2000}
{Zurita} A., {Rozas} M., {Beckman} J.~E., 2000, \aap, 363, 9

\end{thebibliography}
\end{document}